\documentclass{article}

\usepackage{arxiv}

\usepackage{fancyhdr}  

\usepackage{adjustbox}

\usepackage{cite}
\usepackage{amsmath,amssymb,amsfonts}
\usepackage{algorithmic}
\usepackage{graphicx}
\usepackage{textcomp}
\usepackage{subfigure}
\usepackage{amsmath} 
\DeclareMathOperator*{\argmax}{arg\,max} 
\usepackage{algorithm,algorithmic}
\usepackage{url}
\usepackage{amsthm}
\usepackage{subfigure}
\usepackage{multirow}
\usepackage{rotating}
\usepackage{graphicx}
\usepackage{tabularx}
\usepackage{array}
\usepackage{anyfontsize}
\usepackage{color,soul}
\usepackage{graphicx,dblfloatfix}
\usepackage{epstopdf}
\usepackage{blindtext}
\usepackage{amsthm,amssymb,amsmath,bm}
\usepackage{subfigure}
\usepackage{amsfonts}
\usepackage{epsfig}
\usepackage{amssymb}
\usepackage{amsmath}
\usepackage{cite}
\usepackage{graphicx}
\usepackage{multirow}
\usepackage{fancyhdr}
\usepackage{subfigure}
\usepackage[subfigure]{tocloft}
\usepackage[font={small}]{caption}
\usepackage{subfigure}
\usepackage{tabularx}
\usepackage{cite}
\usepackage{adjustbox}
\usepackage{booktabs,multirow}
\usepackage{tabu}
\usepackage{longtable}
\usepackage[table]{xcolor}

\usepackage{datetime2} 
\usepackage{hyperref} 

\title{Leader Selection and Follower Association for UE-centric Distributed Learning in Future Wireless Networks}

\author{
Saeedeh Parsaeefard$^{1}$, Sabine Roessel$^{2}$, Anousheh Gholami Ghavamabad$^{1}$, Robert Zaus$^{2}$, Bernhard Raaf$^{2}$ \\
Corresponding author  e-mail: s\_parsaeifard@apple.com \thanks{This work has been partially supported by the European Commission through the SNS JU project 6G-SHINE (Grant Agreement No. 101095738). 
We thank Ayman Naguib, Mona Chitnis, and Kunal Talwar for their contributions that greatly improved the manuscript. }\\
1: Apple Inc. USA\\
2:Apple Technology Engineering B.V. and Co. KG, Germany  }

\begin{document}
\maketitle




\begin{abstract}
User equipment (UE) devices with high compute performance acting on data with dynamic and stochastic nature to train Artificial Intelligence/Machine Learning (AI/ML) models call for real-time, agile distributed machine learning (DL) algorithms. Consequently, we focus on UE-centric DL algorithms where UEs initiate requests to adapt AI/ML models for better performance, e.g., locally refined AI/ML models 
among a set of headsets or smartphones. This new setup requires selecting a set of UEs as aggregators (here called leaders) and another set as 
followers, where all UEs update their models based on their local data, and followers share theirs with leaders for aggregation. 
From a networking perspective, the first question is how to select leaders and associate followers efficiently. 
This results in a high dimensional mixed integer programming problem and involves internal UE state information and state information among UEs, 
called external state information in this paper. To address this challenge, we introduce two new indices: a Leader Internal Index (LII), 
which is a function of the internal states of each device, demonstrating the willingness to be a leader such as battery life and AI/hardware accelerators, 
and a Leader eXternal Index (LXI), which is a function of external state information among UEs, such as trust, channel condition, 
and any aspect relevant for associating a follower with a leader. These two indices transform the highly complex leader selection and follower association 
problem into a better tractable formulation. More importantly, LIIs and LXIs allow to keep the internal and external state information of this problem 
inside of each device without compromising users' privacy. 
We present the required constraints and objective function, propose a distributed algorithm, and discuss feasibility, challenges, and 
implementation aspects in 5G and 6G. Through simulation results, we demonstrate the performance of our distributed solution compared to the optimal one. 

\end{abstract}

\keywords{Distributed machine learning (DL) algorithms, Fine-tuning, Federated Learning, Subnetworks in 6G, Private networks in 6G}



\section{Introduction}
\subsection{Motivation and Novelties }\label{Motivation}
The next generation of wireless networks will significantly expand the support for distributed machine learning (DL) algorithms 
for both over-the-top use cases and internal networking procedures in training and fine-tuning of models 
through the concepts of federated learning and transfer learning \cite{10056683, 10258360}.  Here, the underlying assumption is that there is an entity inside or outside the wireless network, 
called a leader, and a set of end-user equipments (UEs), called followers. The leader will initiate the algorithm, and until convergence, 
per each epoch, the followers will update their models based on their local dataset and share them with the leader, where the 
leader sends back the aggregated outcome of the followers' models. While this model has the advantage of privacy and reducing 
the overhead of data offloading, it can suffer performance degradation from its lagged response to a data/concept drift where the input data or the underlying 
stochastic relationships between input features and outputs change over time or across different domains due to the dynamic nature of the involved 
environment and behavior of end-users  and wireless networks  \cite{9352556,5288526}. 

In wireless networks, there are diverse sources of noise and stochastic features causing data and concept drift in AI/ML tasks, 
for example: 1) wireless channel variations, such as moving from urban areas to high traffic hotspots in downtowns, 
2) changing UE traffic patterns according to 
specific events, 
3) any personalized behavior of users such as mobility patterns of teenagers during AR/XR sessions, and 4) new location 
or cultural based features in UEs' traffic patterns. 
These stochastic changes lead to direct performance degradation, which can be measured at UE in real-time in contrast to the cloud or inside the network.  
Therefore, to overcome this type of performance degradation, UEs should be able to request a fine-tuning 
of the ML models in a more real-time manner. Along with considering recent UEs, highly equipped with computational and wireless transmission capabilities, 
future wireless networks can introduce UE-centric DL algorithms where UEs can initiate the request for distributed learning and select the leader entities for aggregation of models among themselves, 
without any third party or another network entity. 

This can open a new opportunity for privacy-preserving, personalization, and real-time fine-tuning of ML models, 
and it is aligned with the concept of private networking over future wireless networks for AR/XR use cases. 
To run such a UE-centric DL algorithm, the first question is how to select leaders and followers reliably based on 
UEs' capabilities without sharing UEs' private and internal state information. Let's call it the \textit{``leader selection and follower association problem.''} 
Selecting a leader and associating the followers are related to diverse constraints and mixed-integer variables. 

Due to the diverse nature of these variables, it is not straightforward to present the leader selection and follower association as a tractable optimization 
problem formulation due to non-convex intertwined integer and continuous variables in different dimensions. Also, these internal and external states involve a 
lot of UEs' information, which has privacy and security concerns, and it is not desirable to collect them from UEs. 
Our goal is to keep the related state information inside of UEs. Our goal in this paper is to address the complexity of the leader selection and 
UE association problem to pave the way for real-time UE-centric DL algorithms. First, to provide privacy and not share any UE
state information with other entities, and second, to reduce the complexity of the problem, we introduce two new indices for UEs:
\begin{itemize}
\item Leader  Internal Index (LII) per each UE $n$ ($LII_n$): demonstrating the device's willingness to be a leader. 
This parameter can be defined as an aggregation of the internal states of devices for the parameters that are related to selecting the leader, 
e.g., computation and transmission capabilities, storage capacity, and battery life time.
\item Leader eXternal Index (LXI) between each UE $m$ and UE $n$, denoted as $LXI_{m,n}$: showing the willingness of UE $m$ to select UE $n$ as 
a leader based on UE $m$ state information related to UE $n$, such as proximity of devices, trust factors, channel condition, and any matching factor 
between any pair of UEs. 
\end{itemize}
From above, LIIs show UEs' internal capacity and capability to act as leaders. However, LXIs show the inter-relation among UEs and their 
vote for one UE to be a leader. UEs can determine both of these two indices. It would not be easy to agree on a detailed formulation of how the 
LIIs and LXIs depend on the various parameters due to privacy reasons. 
However, we assume that each UE or each pair of UEs can answer the following questions: 

\begin{enumerate}
	\item ``on a scale from 0 to 10, how would you estimate your capabilities and your willingness to become a leader?” and 
	\item ``on a scale from 0 to 10, how would you estimate your capabilities and your willingness to select UE $n$ as leader?”. 
\end{enumerate}

Here, we assume that $LII_n=0$ means UE $n$ is not willing to be a leader, and $LII_n=10$ means the maximum willingness and readiness of UE $n$ to be a leader. 
Similarly, $LXI_{m,n}=0$ means there is no willingness for UE $m$ to select UE $n$ as leader based on its own preferences and state information, 
and $LXI_{m,n}=10$ means UE $m$ has a strong preference to select UE $n$ as leader.

LIIs and LXIs can help to formulate leader selection and follower association in a more tractable way where we first introduce a set of 
constraints that can mathematically explain the limitations of this optimization problem as follows: C1: UE $n$ can be either follower or leader. C2: Each leader 
should have at least one follower. C3: UEs with LIIs not exceeding a specific threshold should not be considered as leaders. 
This set of constraints can avoid loops among UEs, e.g., one leader be a follower of another leader. 
We introduce a linear utility function which promotes to select leaders with largest values of LIIs and assigns followers to leaders with 
the largest values of LXI. 
The problem of joint leader selection and follower association (abbreviated: \textit{joint problem}) belongs to the general assignment problems (GAP) which are NP hard.

To overcome complexity, distributed solutions are always appealing due to their scalability and adaptability to a large number of UEs and to the 
highly dynamic nature of wireless networks. 
Our distributed model is based on a sub-optimal two-step formulation of the joint problem, where we first divide UEs into two sets, a set of 
leaders and a set of follower UEs, according to their LII values, then we provide a distributed mechanism where, without the need to share LXIs, 
follower UEs can select their leaders. We also provide mechanisms to deal with marginal scenarios, including leaders without followers and leaders with limited connections, based on offering edge servers as an option for isolated UEs. Finally, we demonstrate that our approach can be adapted for 5G and 6G via a related signaling procedure. 

\subsection{Literature Review }
This paper resides at the intersection of architecture aspects of DL algorithm on the edge of wireless networks and the proposed solutions to handle 
complexity of the assignment problems. The concept of federated learning (FL) \cite{mcmahan2017communication} is the main architecture view for DL algorithms 
across decentralized agents (e.g., UEs) while preserving data privacy. 
In traditional FL, a single coordinating server aggregates updates from multiple client devices. To adapt FL for 5G and 6G with 
growing interest to utilize edge resources as aggregators of models, hierarchical FL (HFL) architectures have been proposed in this context, e.g., 
\cite{HFL,FSM, MAFL, 10423299}, where multiple servers at the edge aggregate the models  with the goal of reducing the 
communication overhead compared to that of single server FL. The main assumption in \cite{FSM, MAFL, 10423299} is that the edge servers (aggregators) are 
part of the network in a predetermined location. However, in our UE-centric approach, a subset of UEs, namely leaders, can perform model aggregation to 
introduce more flexibility among UEs. Moreover, UEs can coordinate DL algorithm and the selection of leaders and UE assignment through a distributed 
algorithm. According to our best knowledge, such UE-centric version of HFL algorithms is not studied and our proposed solution can fill this gap. 

In parallel, fully decentralized FL frameworks without a central server have been extensively investigated, e.g., 
\cite{lalitha2018fully, gholami2022trusted, liu2022decentralized}. Here, each UE shares its local model update with its one-hop neighbors at each round of 
communication. Then, UEs implement consensus techniques followed by local training per each UE performing aggregation at all UEs, until convergence. 
Clearly, this model requires extra signaling overhead compared to server-based FL. 
To overcome this challenge, we assume only UEs selected as leaders will aggregate models received from their follower UEs. 
Such an assumption aids in reducing the communication overhead and improving scalability. 

From the solution perspective, our proposed formulation for leader selection and follower association shares similarities with the classic 
combinatorial optimization problems such as the generalized assignment problem (GAP) and facility location problem (FLP); however, it diverges in several key aspects. 
The GAP seeks the minimum-cost or maximum-profit assignment of a number of tasks to a set of servers where each task is assigned to precisely one server agent, 
subject to servers' capacity restrictions \cite{cattrysse1992survey}. The FLP consists of a set of potential facility sites and a set of demand points to be served 
by the facilities. The objective of FLP is to open a subset of facilities and assign demand points to them while minimizing the total cost of opening facilities and 
demand points assignment \cite{kunalpaper}. Another variant of FLP is the capacitated FLP (CFLP) assuming capacity constrains for the facilities, 
that is more relevant to the leader selection and follower association problem. While the set of servers and facilities are known in GAP and CFLP, 
in our problem, each UE can play the role of either a leader (facility/server) or follower (demand point) based on the assignment costs that vary with UEs' $LII_n$ and $LXI_{m,n}$. 
Consequently, we have extra constraints in order to jointly determine the leaders and followers while avoiding the situations of having leaders without followers. 
Therefore, the set of constraints in our proposed problem is more complex which shrinks the feasibility region. 
To cover this point, we propose mechanisms to escape from infeasibility such as introducing an incentive mechanism and having an 
edge server as a backup leader. Still our focus in this paper is not to solve the joint problem more efficiently than existing approaches, 
since we believe there exists extensive background which can be used here directly. Our main focus is to have an efficient distributed algorithm 
given new definitions of LIIs and LXIs to further enhance the scalability and adaptability of the proposed approach and then, 
to benchmark its performance with the optimal solution of the joint problem. 

Our proposed solution falls into clustering-based solutions investigated in FL, e.g., \cite{9207469, long2023multi}. The objective of these works is to 
enhance the model aggregation mechanism by clustering users using clustering methods, such as hierarchical clustering and k-means in the case of data 
heterogeneity, and training a personalized model per cluster. These clustered FL solutions consider only the data and model aspects in clustering and 
the model aggregation is performed in one aggregator. Our proposed method 
is an extension of such clustered distributed learning algorithms where we consider more deployment aspects for clustering with LIIs and LXIs. 
Also, the definition of our cluster for one leader with its set of followers fits to the concept of sub-networks in 6G \cite{D2.2}. 
We will demonstrate how our approach can be integrated easily into 6G sub-networking and private networking and 
be part of the compute and communication fabric in 6G for more UE-centric distributed scenarios. 
The proposed distributed approach, LIIs, and LXIs can inherently be part of 6G to facilitate real-time fine-tuning in the sub-network.

A leader selection mechanism is also part of many network protocols like, e.g., Apache ZooKeeper \cite{zookeeper}. ZooKeeper offers a 
centralized service for maintaining configuration information for a group of servers supporting distributed applications. In this protocol, 
leader candidates are represented by a list of child znodes ranked according to their sequence number, i.e., according to the order in which these 
znodes were created by the servers. The server associated with the lowest sequence number is chosen as leader. When a leader goes out of service, 
its znode is removed from the list, and the server which created the next znode moves up as new leader. Thus, in contrast to the LIIs used in our solution, 
resource and energy constraints and the internal states of the servers are not taken into account for the ranking of leader candidates. 
Furthermore, ZooKeeper does not consider external influence factors like proximity or trust which can play an essential role for leader selection between UEs 
in a cellular network and which are represented by the LXIs in this paper.

\subsection{Paper Organization }
The organization of this paper is as follows. Section \ref{System_Model} contains system model and definitions of LII and LXIs. In Section \ref{Centralized}, we formulate the joint leader selection and follower association   problem, introducing a set of constraints, followed by a centralized solution, feasibility study, and computation complexity. Section \ref{Distributed} includes the distributed algorithm and discussions of marginal scenarios. In Section~\ref{Architecture}, we discuss how our UE-centric algorithms can be adapted for 5G and 6G architectures. Simulation results are presented in Section \ref{Simulations}, followed by conclusions in Section~\ref{conclusions}. 

\section{System Model}\label{System_Model}

\begin{figure}
\begin{center}
\includegraphics[width=3.3 in]{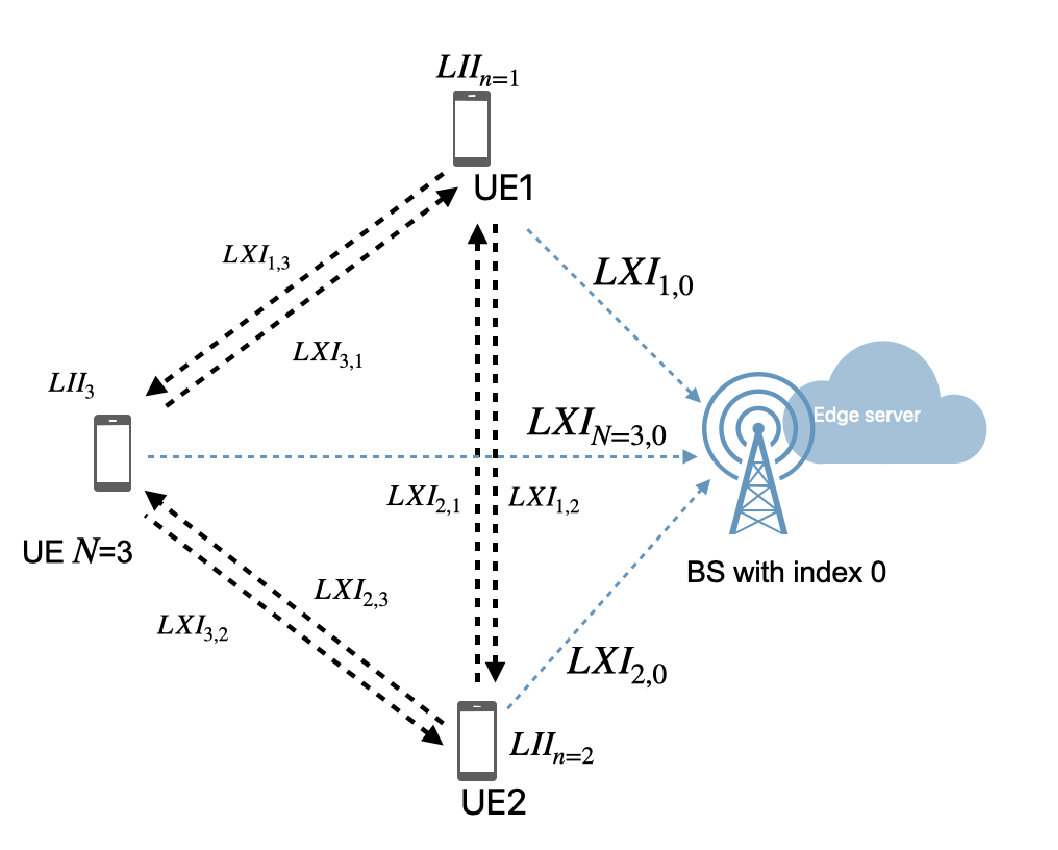}
\caption{ System model of $N=3$ UEs in the coverage of a base station (BS) equipped with edge server, denoted by index $0$}
\label{systemmodel}
\end{center}
\end{figure}

Assume the coverage area of a local hot spot, e.g., airport or stadium, with a set of devices denoted by $\mathcal{N}=\{1, \cdots, N\}$ as depicted 
in Fig. \ref{systemmodel}. We denote the BS with index $0$ assuming only one BS in the coverage area\footnote{Easily our algorithms can be extended to a 
multi BS scenario.}. 

Assume UEs in our setup have specific pre-trained AI/ML models which are compatible. 
However, due to stochastic drift of data and context over time and location, the performance may degrade until one of these UEs 
initiates a request for fine-tuning or re-tuning of models jointly with other UEs in the same location. 
First, let's assume that UEs are willing to select leaders among themselves and to share their models with at least some other UEs, 
while not being willing to share with any third parties including application servers. UEs have diverse internal and external states which should be 
considered in leader selection or follower association, for example:  
\begin{itemize}\setlength{\itemindent}{0.8cm} 
\item [\bfseries Class 1:] Storage and computation capabilities such as availability of an AI accelerator in UE
\item [\bfseries Class 2:] Transmission features such as supported frequency bands, number of supported simultaneous connections, and battery level 
\item [\bfseries Class 3:] Channel state information $h_{m,n}$ between UE $m$ and UE $n$, proximity of UEs, and trust level among UEs 
\item [\bfseries Class 4:] UEs AI/ML model compatibility for specific learning and fine-tuning tasks 
\end{itemize} 
According to the above features, we now define the following two indices: 

\begin{itemize}
\item Leader Internal Index $LII_{n}$ for UE $n$, which is a function mapping various parameters and properties of the UE $n$ on Class 1 and Class 2 attributing to the final value of $LII_{n}$. For example, for battery life, UEs can select increasing function $f_n^{i} (x_n^i)$ where $x_n^i$ is battery life state. Calculating LIIs from attributing parameters can be as simple as a weighted sum of relevant parameters in Class 1 and Class 2.  
When $LII_n > LII_m$, it means that UE $n$ has a better internal state to be a leader compared to UE $m$, e.g., UE $n$ has longer battery life than  UE $m$. 
\item Leader eXternal Index $LXI_{m,n}$ from UE $m$ to UE $n$ which is a suitable function expressing trust and connectivity that may depend on Class 3 and Class 4 parameters, e.g., $h_{m,n}$, the  proximity or channel condition between UE $m$ and UE $n$, their  trust level, match of versions of AI/ML models. Again, $LXI_{m,n} > LXI_{m, n'}$ means UE $m$ prefers to be a follower of UE $n$ compared to UE $n'$. 
\end{itemize}
Details of those functions are left for specific applications, we just note higher $LII_n$ respectively $LXI_{m,n}$ indicate better fitness of UE $n$ to act as leader respectively being accepted as leader by UE $m$. Without loss of generality, we consider the normalization $LII_{n}, LXI_{m,n} \in [0,10]$ where 0 means the lowest and 10 the highest score. 

\section{Joint Leader Selection and Follower Association Formulation}\label{Centralized}

To formulate this problem, we define two variables as follows:  
\begin{itemize}
\item Leader selection variable as $y_n\in \{0,1\}$. If $y_n=1$, UE $n$ is a leader, and otherwise follower; 
\item UE association variable as $x_{m,n}\in\{0,1\}$. If $x_{m,n}=1$, UE $m$ is assigned to UE $n$ as follower, and otherwise, $x_{m,n}=0$. 
\end{itemize}
For preventing loops and overlaps among leaders and followers, we propose the following constrains: 
\begin{itemize}
\item UE $n$ can be either leader or be a follower of one other UE, represented as $$C1: \,\,\,\,\, y_n + \sum_{m =1, m \neq n}^{m=N}x_{n,m}=1, \forall n.$$\item Each leader should have at least one follower and UEs not selected as leaders should not have followers:
$$C2: \,\,\,y_n \leq \sum_{m =1, m\neq n }^{N}x_{m, n} \leq (N-1) \times y_n.$$ This constraint forces UEs,  not selected as leaders, not to  have any followers because both sides of C2 will be zero if $y_n=0$. When $y_n=1$, this leader should have at least one follower according to the lower bound of C2 and can have up to $N-1$ followers based on the upper bound. 
\item Only UE $n$ with $LII_n$ larger than a specific threshold $\rho$ should be selected as leader, i.e., $$C3:  \,\,\,y_n \leq \big\lceil {\max(0,LII_n-\rho)} \big\rceil .$$ 
$\rho$ indicates which level of $LII_n$ is required for a UE to be selected as leader; for $\rho=0$ any UE with $LII_n > 0$ can be selected. 
C3 forces UEs with $LII_n \le \rho$ to have $y_n=0$.   \end{itemize}

The objective is to assign UEs to the most eligible leaders with highest values of LXIs. There are many ways to define objective functions for this problem. To get a tractable formulation with reasonable complexity, our goal is to introduce specific linear or semi-definite objective functions. 
For the linear objective function, we can introduce a linear utility function as 
\begin{align}
\sum_{n \in \mathcal{N}}  LII_n y_n + \sum_{n \in \mathcal{N}} \ \  \sum_{m=1, m \neq n}^{N} LXI_{m,n} x_{m,n}  ,  \nonumber 
\end{align}
where the first term is related to the leader selection, promoting leaders with higher values of $LII_n$ and the second term is related to UE assignment, promoting UEs to be assigned to the leader with higher values of $LXI_{m,n}$. Consequently, the following problem can be introduced
\begin{align}\label{eq1}
\max_{\stackrel{y_n, x_{m,n}}{\forall n , m \in \mathcal{N}} } \ \  &\sum_{n \in \mathcal{N}} \left[LII_n y_n + \sum_{m=1, m \neq n}^{N} LXI_{m,n} x_{m,n}\right] \\ & \text{s.t.:}  \,\,\, C1,\, C2, \, C3. \nonumber
\end{align}

The joint leader selection and follower association problem stated in \eqref{eq1} maps to an integer linear programming (ILP) 
formulation resulting in an NP-hard optimization problem. There exist efficient approximation methods such as relaxation of variables, e.g., 
relaxing  $y_n \in [0,1]$ and $x_{m,n} \in [0,1]$ for all $n$ and $m$, branch-and-bound, or cutting plane methods \cite{convexboyd}, which can 
be exploited to solve or at least approximate the problem. In this paper, we apply exhaustive search to obtain an optimal solution for benchmarking the 
distributed algorithm proposed in Section \ref{Simulations}.

Also, it is important to be aware of conditions under which \eqref{eq1} cannot provide a solution. There are two scenarios where the leader selection and 
follower association problem is not feasible:
\begin{itemize}\setlength{\itemindent}{0.8cm} 
\item[\bfseries Case 1:] If LIIs of all UEs are equal to zero, none of the UEs can act as leader for distributed algorithm. 
\item[\bfseries Case 2:] $LII_m \leq \rho$ and $\sum_{n=1, n \neq m}^{N} LXI_{m,n} LII_{n} =0$ for UE $m$, meaning there is no other UE that can be  leader for UE $m$, either because $LII_n=0$ or $LXI_{m,n}=0$. In this case UE $m$ will be isolated because it can neither find a leader nor be leader itself.
\end{itemize}

To escape infeasibility in general, our proposal is to introduce an edge cloud for user-centric distributed learning to provide higher feasibility for 
fine-tuning initiated by UEs in 6G, as depicted in Figure \ref{systemmodel}.
Our formulation can be easily extended for this case by introducing the edge cloud as Node UE $0$ with $LII_0>0$ and $LXI_{0,n}=0$ for all $n$ to ensure 
that the edge cloud is not assigned as follower to any of the UEs. But we will stay with the above formulation for the rest of the paper.

\section{Distributed scenario}\label{Distributed}
Distributed algorithms are highly appealing due to their scalability, particularly in scenarios where collecting and maintaining a global view of the network is either too expensive or impractical. For instance, a distributed algorithm to solve the FLP is proposed in \cite{kunalpaper}. This approach formulates the FLP as a MILP, relaxes it to a Linear Program (LP), approximates the LP using a distributed primal-dual approach, and then employs a randomized rounding procedure to derive the final integer solution of the original MILP. Solving the joint leader selection and follower association problem introduced in Section~\ref{Centralized} using the distributed primal-dual decomposition similar to \cite{kunalpaper } would require extensive signaling among UEs \cite{convexboyd,saeedehbook} since UEs do not have access to all LXIs and LIIs of other UEs.
Consequently, to propose a distributed approach with minimal signaling, we introduce a leader set and a follower set based on the values of $LII_n$ and $\rho$, where 
each UE $n$ can fall into one of these sets:
\begin{itemize}
\item Leader set $\mathcal{L}$ containing all $UE_n$ with $LII_n > \rho$
\item Follower set $\mathcal{F}$ containing all $UE_n$ with $LII_n \leq \rho$
\end{itemize}
Now, \eqref{eq1} is transformed into the follower association problem as follows: 
\begin{align}\label{eqd1}
\max_{\stackrel{x_{m,n},}{\forall n \in \mathcal{L}  , m \in  \mathcal{F}} } \ \  &\sum_{n \in \mathcal{L}} \left[ LII_n  + \sum_{m \in \mathcal{F}} LXI_{m,n} x_{m,n}\right] 
\end{align}
where C1-C3 are dropped from \eqref{eqd1} to reduce signaling. Consequently, \eqref{eqd1} can be decomposed  among followers as: 
\begin{align}\label{eqd2}
\max_{\stackrel{x_{m,n},}{\forall n\in \mathcal{L}}  }\ \  &\sum_{n \in \mathcal{L}} \left[ LII_n  +  LXI_{m,n} x_{m,n}\right]  , \ \forall m \in \mathcal{F}.  
\end{align}

\textit{Remark 1:} Note that separating the leader and follower sets for \eqref{eqd1} and \eqref{eqd2} establishes a sub-optimal approach to the joint problem. 
Creating these two sets based on $\rho$ allows for removing constraints C1 and C3.
However, the separation into disjoint leader and follower sets does not resolve constraint C2 as it cannot be ensured that the there exists a UE~$m$ with 
good level of $LXI_{m,n} $ which follows the leader UE $n$ with $y_n = 1$.
As the proposed distributed algorithm relaxes C2, we introduce a mechanism with two phases to meet the constraint C2.

In our distributed algorithm, at the first phase, we assume that all UEs send their values of $LII_n$ to each other and they will agree on the value 
of $\rho$, the threshold for selecting leaders based on $LII_n$\footnote{For sending these values, UEs can use cellular or peer-to-peer communication.}. 
The value of this threshold can be adjusted among UEs when they will announce their $LII_{n}$. In the distributed case, UE $n$ can set $LII_{n}=0 $ if it is not willing to be part of the leader list. 
Otherwise, if $LII_{n} > \rho$, it should act as leader if UE $n$ receives the request from other UEs. 
Then, accordingly, each UE can belong to the leader set or the follower set: 
\begin{itemize} \setlength{\itemindent}{0.4cm} 
	\item If UE $m \in \mathcal{N}$ belongs to the follower set, i.e., $LII_m \leq \rho$, UE $m$ will calculate $LI_{m,n}=LII_{n}+{LXI_{m,n}}$ and then UE $n^*$ will be the leader of UE $m$ when $n^*= \argmax_{n} LI_{m,n}$ which is a simple heuristic for \eqref{eqd2}. 
\item If UE $n \in \mathcal{N}$ belongs to leader set, i.e., $LII_n > \rho$, it should wait to receive requests from other UEs, i.e., Step 1-4 at Phase 1. 
\end{itemize}
At Phase 1, the above proposed approach ensures that each UE can be either leader or follower, i.e., C1, and the leader with the 
best $LII_n$ and $LXI_{m,n}$ will be selected which can maximize the objective function. Also, only UEs with $LII_n > \rho$ can be selected as leader, i.e. C3 
also holds thanks to the above algorithm. However, at Phase 1, we cannot ensure that C2 holds, i.e., each leader has a follower. 
To handle this point, we introduce a timer threshold $T$ that will be executed at Phase 1. This timer allows all leaders and followers to setup 
their clusters during a window of $T$ in Phase 1. If a UE belonging to the leader list did not get any follower requests, it means it is an isolated leader 
(i.e., without follower) and C2 does not hold. Then, after $T$, Phase~2 will be initiated where this leader will act as follower, i.e., select its leader 
and sends a request to it. 
To facilitate building the leader list for these remaining UEs, we introduce the following two steps: 

\begin{itemize} 
\item \textbf{Step 2-1:} After $T$, leaders who have followers, will announce their $LII_n$ for the next round. 
\item  \textbf{Step 2-2:} UEs belonging to $\mathcal{L}$ from first round, which do not have a single follower, are moved to $\mathcal{F}$ and will select their 
leaders according to $n^*= \argmax_{n} LI_{m,n}$ and send request
s to their respective leaders $n^*$. 
\end{itemize}
With these two rounds, C1, C2, and C3 can be satisfied in a distributed way and the leaders with maximum values of $LII_n$ and $LXI_{m,n}$ will be selected, for all $n \in \mathcal{L}$ and $m \in \mathcal{F}$. 
If there is UE $m$ with $LXI_{m,n}=0$ for all $n \in \mathcal{N}$, it means UE $m$ does not want to accept any other UE as leader for DL algorithm\footnote{We assume all the nodes can be inside of the coverage area of cellular networks or leaders. Considering hidden node problem or fading and low coverage  is left for future investigation. }.

\begin{algorithm}
\caption{Distributed Algorithm with 2 Phases}
\label{alg1}
\begin{algorithmic}
	\STATE \textbf{Initialization:} For all $n \in \mathcal{N}$, calculate $LII_n$. Set $T$ and $\rho$. If $LII_n>\rho$ announce it to the peers. 
	\STATE \textbf{Phase 1}:
	\begin{itemize}
		\item \textbf{Step 1-1:} All UEs will calculate $\mathcal{L}$ and $\mathcal{F}$
		\item \textbf{Step 1-2:} For all UE $m$ in $\mathcal{F}$, calculate $LXI_{m,n}$ and $LI_{m,n}=LII_{n}+{LXI_{m,n}}$ for all $n \in \mathcal{L}$
		\item \textbf{Step 1-3:} UE $m$ will send the request to $n^*$ from (1-2) 
		\item \textbf{Step 1-4:} UE $n$ should send ACK to all requests from all other UEs during $T$
	\end{itemize}
	\STATE \textbf{Phase~2}:
	\begin{itemize}
		\item \textbf{Step 2-1:} At the end of time $T$, UEs with followers will announce their LIIs again 
		\item  \textbf{Step 2-2:} Leaders without followers act as followers, and run Steps 1-2 through 1-3 to send request to their selected leaders, to be acknowledged as in Step 1-4
	\end{itemize}
	\STATE \textbf{Terminating the distributed algorithm :} The leader selection and follower association terminates after Phase~2 and DL algorithm is initiated. Those UEs belonging to follower list with $LXI_{m,n}=0$ are out of the distributed algorithm, i.e., cannot participate. 
\end{algorithmic}
\end{algorithm}

In the distributed algorithm, UEs do not need to share their LXIs, and UEs will send the request to the leaders based on their LXIs and their list of leaders. 
All leaders in $\mathcal{L}$ must ACK the requests from followers. The distributed algorithm can always converge thanks to the second phase, where leaders without followers turn into followers. 
However, achieving a high or even optimal utility value \eqref{eq1} depends on an appropriate selection of $\rho$. 
In the distributed scenarios, there are some marginal scenario (cases) as follows:
\begin{itemize}\setlength{\itemindent}{1.4cm} 
\item[\bfseries Scenario 1:]
Given $\rho$ in Phase 1 of distributed algorithm,  for all $n \in \mathcal{N}$, we have $LII_n > \rho$ , meaning that all UEs belong to the candidate  leader  set and there is no UE in the follower set to send a request to leaders. In this case, in Phase~2, all the leaders act as followers and ask for the leader but there is no leader in the leader set of Phase~2. Consequently, there is no leader at the end of the process and the distributed algorithm will be either terminated or each UE should select the edge server as leader, 
to achieve a feasible scenario without a lot of iterations. 
\item[\bfseries Scenario 2:] Given $\rho$ in Phase 1 of the distributed algorithm,  if for all  UE $n \in \mathcal{L}$ with $LII_n > \rho_n$ we have $LXI_{n,m}=0$ for all ${m} \in \mathcal{F}$, i.e., none of UEs in the follower set  trusts any UE in the leader set. 
In this case the UEs can either cancel DL algorithm or resort to the edge server. 
\item[\bfseries Scenario 3:] For all $n \in \mathcal{N}$, $LII_n=0$, meaning no UE is willing be a leader. Here, UEs can either cancel the distributed AI/ML task or request the network to set up a task with a central server, or ask the network to provide an incentive to at least some UEs, thus persuading them to step up and become leaders. 
\end{itemize}

\begin{figure}
\begin{center}
\includegraphics[width=3.1in]{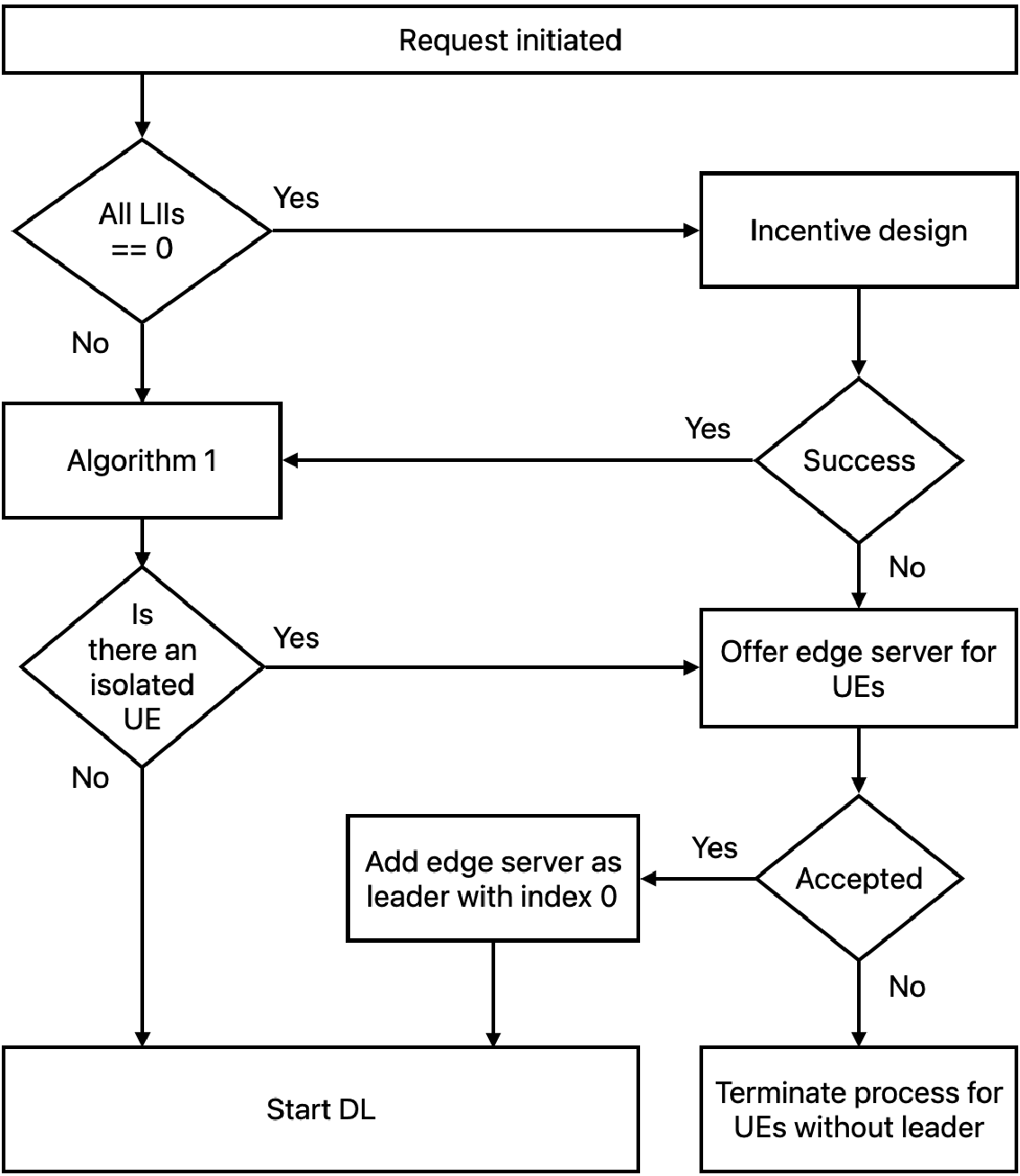}
\caption{ Process of proposing edge server to cover marginal scenarios in distributed algorithm.}
\label{Dia1}
\end{center}
\end{figure}

The process of proposing the edge server to cover marginal scenarios is presented in Fig. \ref{Dia1}. This process is started  by checking if there is an available leader. If there is none, e.g., $LII_n=0$ for all $n$,  the incentive mechanism  is started to encourage UEs to recheck or change their willingness to lead the DL algorithm. If the incentive mechanism is successful,  Algorithm \ref{alg1} is started. It is also started when there is at least one UE with $LII_n \neq 0$. The process will check if there are any isolated UEs. If yes or if the incentive mechanism is not working, then the network will propose the edge server. Then DL algorithm will be started for UEs with leaders (incl. edge server) and will terminate for isolated UEs.  

Our formulation and proposed Algorithm \ref{alg1} can cover the scenario that each leader is limited in some of its resources, e.g., 
computation, storage, or communication. All these resource limitations can be translated to a maximum number of followers which each leader can accept in 
constraint C2. Here, for UE $n$ allowing a maximum number of accepted followers $N_n^{Lim}$, constraint C2 will change to:
$$C2^{Lim}: \,\,\,y_n \leq (\sum_{m =1, m\neq n }^{N}x_{m, n}) \leq N_n^{Lim} \cdot y_n.$$ 

This allows to apply solution methods similar to the ones proposed for the joint problem. 
For the distributed algorithm, however, we need to modify some steps to cover this modification: 
First, each leader should have both acknowledge signal  (ACK) and non-acknowledge signal (NACK) responses. 
Also, by default, we assume all UEs belonging to the follower list will start accessing their optimal leaders. 
Leaders will serve followers first come first serve. 
If a UE receives ACK, it means the leader has enough resources to serve that UE, otherwise the leader doesn't. 
In the latter case, the follower will remove this leader from its list and finds the second most eligible leader in Step 2. 
The remaining steps will be applied for this new leader. The algorithm will continue until all UEs belonging to the follower list have found a leader or their leader list is empty. 
It is clear, that the distributed algorithm still has a convergence point. Reaching the optimal solution will be less likely though, since it is not clear if UEs with the best $LI_{m,n}$ can access to leader $n$. 
These constraints make it more likely that some UEs cannot find a leader. One approach to resolve the above issue of
followers with an empty leader list is to use the edge server as a way to improve the performance 
of the distributed algorithm. UEs can accept or reject the edge server based on their $LXI_{m,0}$.

\section{Architecture Aspects and Signaling for Future Wireless Networks}\label{Architecture}

Our proposed approach can be easily adapted for 5G and 6G. Assume we can introduce a new virtual or physical network function, which we referred to a new network function (NNF) for UE-centric DL algorithm. When one of UEs initiates DL algorithm, this NNF can create the list of UE candidates for this request based on some location based information (such as location area code or cell information) and some user consent information (such as willingness to participate in DL algorithm). Then, NNF can broadcast the request of this new UE-centric DL algorithm to all UEs in its candidate list where these UEs will announce their LIIs and LXIs to the NNF for solving the centralized solution or they can run their decentralized solution where they inform each other about their LIIs.

\begin{figure*}
\begin{center}
\includegraphics[width=6.3 in]{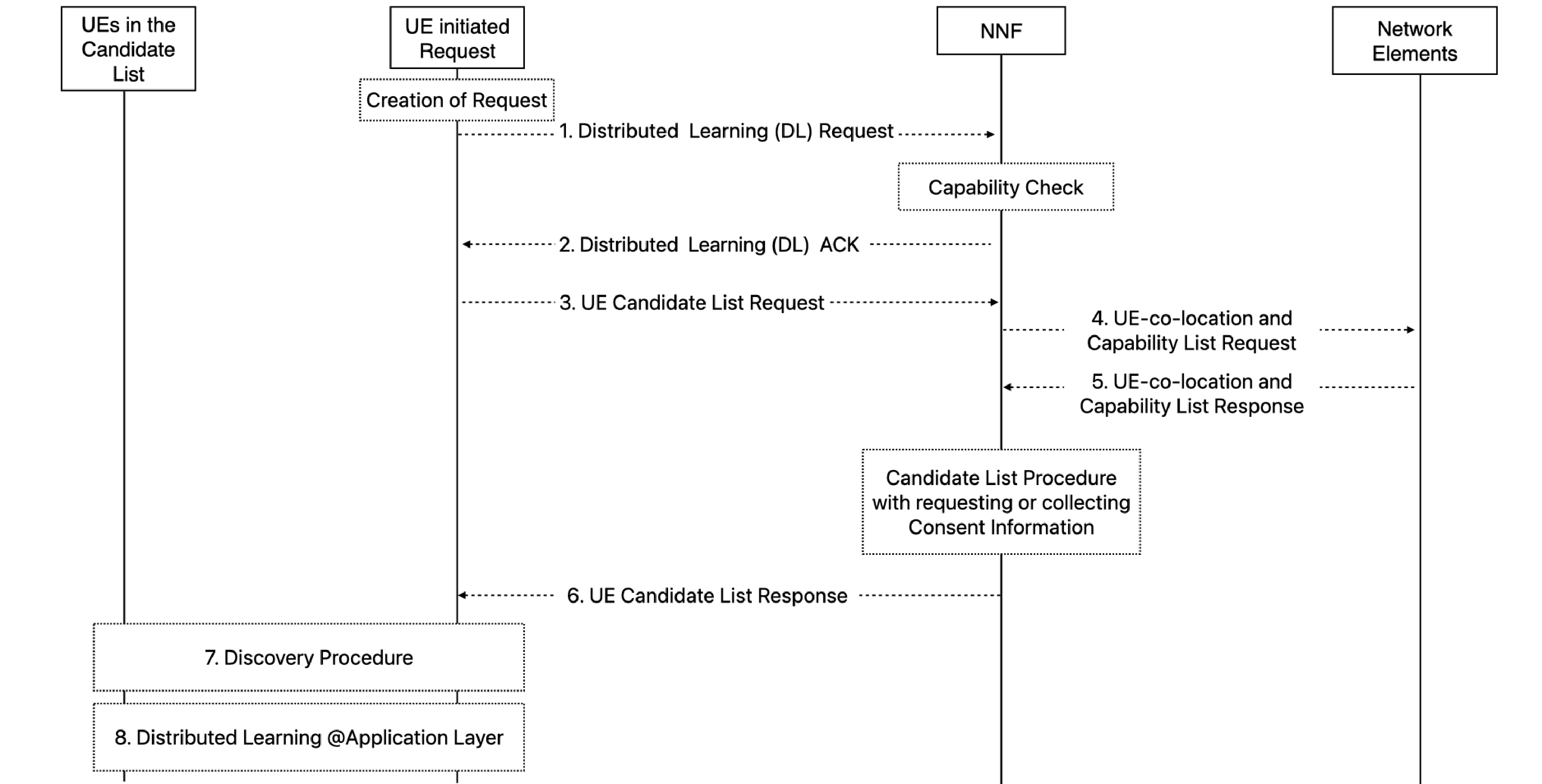}
\caption{The network procedure and signaling for UE-centric DL algorithm in 5G and 6G}
\label{NNF}
\end{center}
\end{figure*}

The signaling related to the above procedure is presented in Figure \ref{NNF}. In Step 1, after initiating the procedure by UE, NNF will get the control message and start the compatibility check procedure. This procedure is responsible to assure the capabilities of this UE for such a request and it will check capabilities and authentication status. For example, in 5G, this compatibility check can be handled with the help of HSS (Home Subscriber Service) of UDM (Unified Data Management) where AMF (Access and Mobility Management Function) can connect NNF to HSS and UDM. After sending  ACK to this UE (Step 2), UE will ask for a candidate list of other UEs, willing to attend this procedure. NNF will create this candidate list which is a list of UEs who are eligible to be part of the process from location perspective (proximity of UEs), compatibility of computation and communication and other related parameters available at network not based on the internal state information of UEs or their private information. Steps~4 and~5 are related to collecting the information by NNF from other core or RAN entities to create the candidate list. Then, NNF will broadcast the request of UE-centric DL algorithm to all UEs in the candidate list (Step 6). Then, in Step 7, the discovery procedure will be started which includes the leader selection and follower assignment algorithm either in centralized or distributed manner according to our proposed methods in Section \ref{Distributed}. In the final step (Step~8), the UE-centric DL algorithm will be executed among UEs and it will be terminated after convergence. NNF can be hosted by one of the core network functions, e.g., AMF and NWDAF (Network Data Analytics Function), or be part of cloud RAN, e.g., RIC (RAN Intelligent Controller), depending on how the network will be adapted for this procedure. The main difference between 5G and 6G for this procedure is about the location of NNF and how to handle steps~4 and~5 for collecting information. In 6G, the signaling to collect the required information can be optimized according to the location where NNF will be hosted, while for 5G, we can use the existing signaling to collect this information. Note that adding the required signaling collecting the values of LIIs and LXIs for the centralized solution is straightforward for both 5G and 6G.  

\section{Simulation Setup and Results}\label{Simulations}
Through the simulation results, we will present the performance of our distributed algorithm compared to an optimal solution to the joint problem found by exhaustive search. 
We will compare in terms of utility, complexity and computation time for different scenarios such as number of UEs $N$ and values of $\rho$, giving different sizes $L := |\mathcal{L}|$ of the candidate leader set $\mathcal{L}$, respectively the follower set $F := |\mathcal{F}|$.

\subsection{ Model for LII and LXI}
For a simple model of LIIs and LXIs, we assume that the participating UEs have sufficient compute capability and describe the UE internal state $LII_n$ by its 
battery state $LII_n \in \{0,1,\cdots,10\}$ where LII values are set based on a uniform distribution.
We model the UE external state as a matrix $LXI=[LXI_{m,n}]_{\forall m, n }$, with zeros in the diagonal and which is typically not symmetric.
This means that the external state is uni-directional, or in other words, the $LXI_{m,n}$ describing the external state of UE $m$ toward UE $n$ can be 
different from $LXI_{n,m}$ of UE $n$ toward UE $m$. This may be due to differences in trust levels or in channel gains. Again $LXI_{m,n}$ values are set based on a uniform distribution such that $LXI_{m,n} \in \{0,1,\cdots,10\}$.
The distributed algorithm obtains a set of candidate leaders $\mathcal{L}$ by setting the threshold $\rho>0$. A rule for choosing $\rho$ needs to be provided or agreed upon. 
One rule could be that each UE sets $\rho = \mu(\{LII_{n}\})$, where $\mu(\cdot)$ refers to the mean value, such that all UE $n$ with $LII_n > \rho$ are part of the candidate leader set.
Another rule could be to select $\rho > 0$ such that $L \leq \lfloor N/2 \rfloor$ for the size $L := |\mathcal{L}|$ of the candidate leader set.
We focus on $N \in \{7,8, \cdots,12\}$ and per each $N$, we generate 100 instances of LIIs and LXIs, expect for $N=12$ where we ran only 30 instances. 
For each instance, we run the distributed algorithm and the exhaustive search to derive  both the utility function and 
the related set of leaders and their associated followers.

\subsection{Benchmarking the Distributed Algorithm}

\begin{table*}[t]\footnotesize	
\renewcommand{\arraystretch}{1.4}
\begin{center} 
\begin{tabular}{ | p{0.18 \textwidth}|p{0.36 \textwidth}|p{0.33 \textwidth}|} 
	\hline
	\multirow{2}{*}{Setup} & Distributed Algorithm with $N$ UEs  and \newline  given sets $\mathcal{L}$, $\mathcal{F}$ of candidate leaders, followers      & Exhaustive Search with $N$ UEs and  \newline any subset of $k$ leaders $1\le k \leq \lfloor N/2\rfloor$\\
	\hline 
	Possible configurations & $ < \sum_{k=1 } ^{ \min(L,\lfloor{N/2}\rfloor)}k! \cdot \binom{L}{k} \cdot \genfrac{\{}{\}}{0pt}{}{N-L}{k} \cdot k^{(L-k)}$ & $\sum_{k=1 }^{ \lfloor N/2\rfloor} k! \cdot \binom{N}{k} \cdot \genfrac{\{}{\}}{0pt}{}{N-k}{k} \phantom{\bigg|}$  \\ 
	\hline
	Complexity & P & NP-hard \\
	\hline
	Runtime & $\mathcal{O}(N)$ &$\mathcal{O} \left(\left\lceil\frac{N}{2}\right\rceil ! \cdot 2^N \cdot  N^3\right) \phantom{\bigg|}$  \\
	\hline
\raggedright Number of signaling messages   & $3N + L - 2$ &   $N+1$ \\
	\hline
\end{tabular}
\caption{Comparison of the distributed algorithm with the optimal solution found by exhaustive search.}
\label{Table1}
\end{center}
\end{table*}

We compare the distributed algorithm with the optimal solution found by exhaustive search in Table \ref{Table1} w.r.t.~the number of possible configurations, 
the complexity of the problem, the order of the runtime complexity w.r.t.~$N$, and the overall amount of messages required.
For the parameters provided in the first row of Table \ref{Table1} referred to as setup, the second row captures the number of possible cluster configurations%
\footnote{To ensure that a leader has at least one follower, we find for the $\binom{N}{k}$ possible leader configurations (selection of k leaders from N UEs) and all ways to partition a set of $n$ UEs (followers) 
	into $k$ non-empty subsets.
	The number of ways to partition accordingly is the Stirling number of the 2nd kind which is denoted as $\genfrac{\{}{\}}{0pt}{}{n}{k}$ \cite{Knuth_2Notes}.
	Subsequently, there are $k!$ ways to associate these $k$ subsets with the $k$ leaders. 
	For the joint problem $n = N-k$. For the distributed algorithm, we first partition $n = N-L$ followers and finally also assign each of the $L-k$ isolated leaders to one of the $k$ subsets, giving a factor of $k^{(L-k)}$. 
	For both cases we then sum over the relevant $k$.}.%
The upper bound to the distributed algorithm's number of cluster configurations is indeed smaller than the number of cluster configurations to be visited by exhaustive search. To compare both expressions, 
we use the following rule for Stirling numbers of the 2nd kind: $\genfrac{\{}{\}}{0pt}{}{n+1}{k} = k\genfrac{\{}{\}}{0pt}{}{n}{k} + \genfrac{\{}{\}}{0pt}{}{n}{k-1} \geq k\genfrac{\{}{\}}{0pt}{}{n}{k}$ and 
show for each summand, comparing the exhaustive search summand with the distributed algorithm summand, that $\genfrac{\{}{\}}{0pt}{}{N-k}{k} \geq k^{(L-k)}\genfrac{\{}{\}}{0pt}{}{N-L}{k}$. 
For regular scenarios where $L < N$, the number of possible cluster configurations of the distributed algorithm is strictly smaller than the one for the exhaustive search. 
The third and fourth rows present the complexity and the order of the runtime, with respect to $N$, of the distributed algorithm as compared to the exhaustive search algorithm.
While the distributed algorithm's complexity is clearly an advantage, the cost of the 
distributed algorithm is the overhead of the extra signaling which needs to be passed among UEs. 
Our analysis gives an upper bound of the overall amount of messages required in the fifth row of Table \ref{Table1}. 
For the centralized solution, $N$ UEs will send one message including $LII_n$ and $LXI_{n,m}$. 
One broadcast message can inform all UEs about their configuration, as we explained in Fig. \ref{NNF}. 
Therefore, the number of messages is $N+1$. In Algorithm \ref{alg1}  
all UEs should inform each other about their LIIs. 
If UEs can use a broadcasting channel (as we explained in Section \ref{Architecture}), this step requires $N$ broadcasting messages among UEs. Then, in Phase 1 of Algorithm \ref{alg1}, at least all followers should send their request to their selected leader and receive an acknowledgment signal, which results in $2F$ messages.  Phase~2 only requires messaging between the isolated leaders and the leaders with followers, i.e., leaders with followers announce their LIIs and the isolated leaders will respond with a follower request. Therefore, signaling in Phase~2 for announcing LIIs is in the worst case $L$  and $2(L-1)$ for request and ACK among isolated leaders and leaders with followers (upper bound by assuming there is only one leader with followers from Phase 1), respectively. The total number of messages in this cases is equal to  $N + 2F + L + 2(L-1)$ which can be simplified as $ 3N + L - 2$.\footnote{If UEs need to use peer-to-peer communication, Phase 1 requires $N(N-1)$ messages among UEs at worst case in full mesh scenario which can be reduced to 2(N-1) with hierarchical peer-to-peer connection. The number of messages for signaling in Phase~2 for announcing LIIs in the worst case (assuming that there is no isolated leader) is $L(L-1)$ for peer-to-peer communication and $2(L-1)$ for request and ACK among isolated leaders and leaders with followers (upper bound by assuming there is only one leader with followers from Phase 1), respectively. The worst case messages among UEs in that case is $N(N-1)+ 2F +L(L-1)+2(L-1)$ which can be simplified to $N(N+1)+L(L-1)-2$ }  
While our approach adds extra messaging among the UEs compared to the centralized approach, still the number is tolerable compared to other algorithms such as the
primal-dual approach stated in \cite{kunalpaper}. 
Our approach requires one round of messaging of $3N + L - 2$ effort ($\mathcal{O}(N^2)$ for peer-to-peer), while for primal-dual approaches each iteration
needs the UEs to send their Lagrange variables with $N^2$ messages. 
Different from \cite{kunalpaper}, our proposed method is capable of trading the amount of signaling for a considerably reduced computation complexity.

For finding the optimal solution, it is sufficient to visit all possible leader-first follower configurations, a configuration being a certain permutation of a certain set. Then, we need to calculate the 
utility function for each of the leader-first follower configurations by adding the maximum of $LXI_{m,n}$ entries for the remaining, not yet assigned UEs~$m$ to the utility of the leader-first follower configuration. 
With this explanation of the exhaustive search, we now present our simulation results. 

\subsection{Cluster Configurations}
\begin{figure*}[t]
\begin{center}
\includegraphics[width=6.4 in]{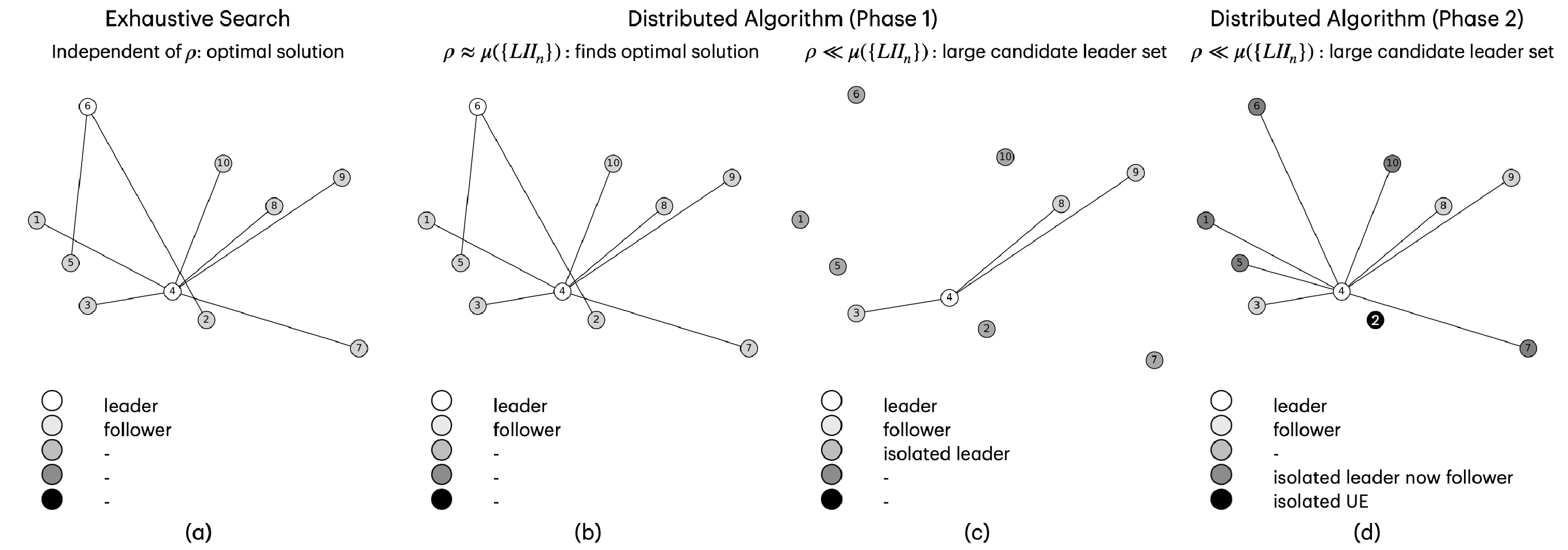}
\caption{Cluster configurations from left to right: 
(a) Exhaustive Search, 
(b) Phase 1 of the Distributed Algorithm for an appropriate value of $\rho$, 
(c) Phase 1 of the Distributed Algorithm for a too small value of $\rho$, 
(d) Phase~2 for the too small value of $\rho$. 
All figures for an identical set of LIIs and LXIs.}
\label{N10Illu}
\end{center}
\end{figure*}
In Fig. \ref{N10Illu}, we compare the cluster configurations created by the distributed algorithm with the ones created by the joint problem's optimal solution.  
Fig. \ref{N10Illu} (a) shows the result of the joint problem for one specific set of LIIs and LXIs where UEs 6 and 4 are selected as leaders, and the remaining UEs are assigned to them as followers based on their LXI values. 
In Figs. \ref{N10Illu} (b) and (c), we have selected two different values of $\rho$. The first value of $\rho$ results in an optimal set of leaders (UE 4 and 6), 
whereas for $\rho \ll \mu$ the set of candidate leaders is very large. 
Hence, we see that in Phase 1 UE 4 is selected as a leader while all other candidate leaders are isolated. As shown in Fig. \ref{N10Illu} (d), all isolated leaders act as new followers and most of them follow UE 4 based on 
their LXI value. 
There are two interesting points which demonstrate the divergence of the distributed algorithm based on the value of $\rho$. 
First, UE 6 does not have a chance to become a leader, and second UE 2 cannot connect to UE 4 as a leader based on its LXI value to UE 4. 
Contrary to the distributed algorithm, the optimal solution found by exhaustive search does not create isolated leaders and selects an optimal leader set (UE 4 and 6). 
However, depending on the LXIs, \textit{isolated UEs} can occur in the optimal solution (Infeasibility Case~2). Finally, in Fig.~\ref{N10Illu} (d) the best leader for UE 2 is not available anymore in Phase~2 because 
UE 6 has turned into a follower.

\begin{figure*}[b]
\begin{center}
\includegraphics[width=\textwidth]{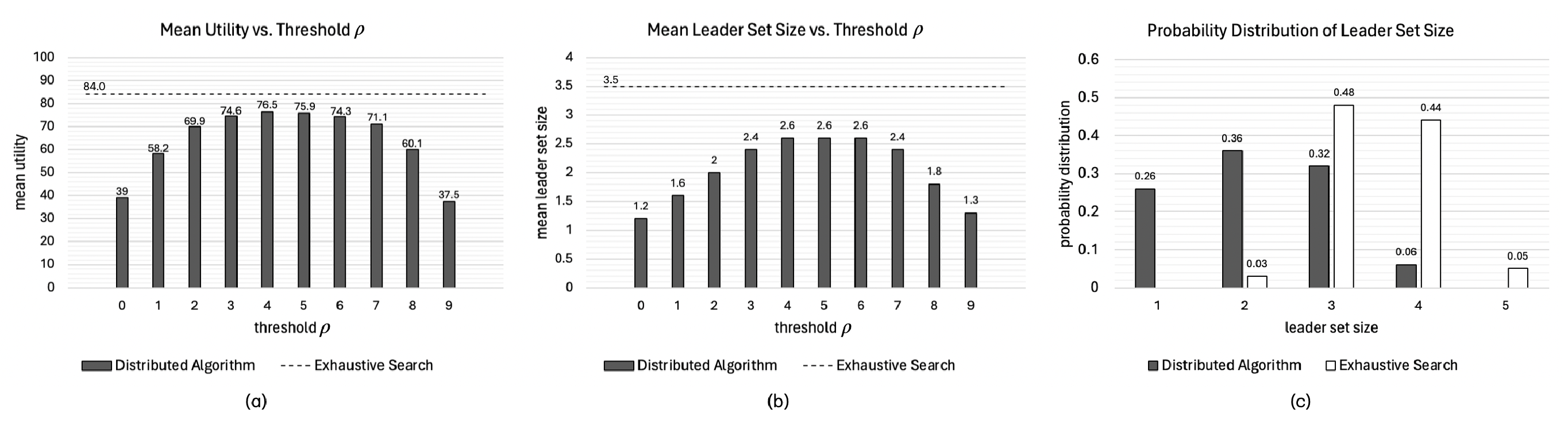}
\caption{Distributed algorithm’s (a) mean utility value and (b) mean (resulting) leader set size across thresholds $\rho \in \{0,1\cdots, 9\}$ for $N=10$ compared to optimal solution’s mean, (c) leader set size probability distribution (average over all $\rho$) of distributed algorithm and of optimal solution.}
\label{N10TR}
\end{center}
\end{figure*}

Figure \ref{N10TR} (a) compares the distributed algorithm with the optimal solution for $N=10$ UEs and $\rho \in \{0,1, \cdots, 9\}$. 
This figure shows that the optimal value for $\rho$ is $\rho \approx \mu(\{LII_n\}) = 5$. At this point, the mean utility of the distributed algorithm 
stays within $10\%$ loss from the optimal solution's mean utility. Very small or very large values of $\rho$ effectively limit the number of available followers (small $\rho$) or the number of available leaders (large $\rho$) 
for the distributed algorithm such that the gap between the distributed algorithm and the optimal solution is enlarged. 
This result highlights that selecting an appropriate value of $\rho$ is essential for the distributed algorithm. 
As mentioned earlier, in Phase~1 of Algorithm \ref{alg1} all UEs would be candidate leaders if $LII_n > \rho$ for all $n$. 
In this case, Algorithm \ref{alg1} terminates with the number of leaders equal to $0$, because at the end of Phase~2 none of the UEs has been selected as leader. 

In Fig. \ref{N10TR} (b), we present the mean leader set sizes for $\rho \in \{0,1, \cdots, 9\}$ for 100 realizations of LIIs and LXIs each, compared to the mean leader set size (here 3.5) of the joint problem's optimal solution.
For very low or large values of $\rho$, the difference between the mean leader set size of the distributed algorithm and the optimal solution is larger 
than for the optimal value of $\rho$, or in other words the chance of picking the best set of leaders is smaller than for the optimal value of $\rho$.
For $\rho$ of value 4, 5, or 6, the mean leader set size of the distributed algorithm approaches the optimal one. 
We also provide the probability distribution functions (PDFs) of the leader set size for the joint problem's optimal solution and for the distributed algorithm in Fig. \ref{N10TR} (c). 
For an evaluation, we fit our data points with a Gaussian distribution where the mean and the variance for leader set size for distributed algorithm are 2.1 and 1.05, 
and for the optimal solution are 3.47 and 0.63, respectively. This evaluation confirms, that in general the optimal solution  
tends to higher leader set sizes and a lower variance as compared to the distributed algorithm trending toward smaller leader set sizes and a higher variance.

\begin{table*}[t]\footnotesize	
\renewcommand{\arraystretch}{1.4}
\begin{center}
\begin{adjustbox}{width=\textwidth}
\begin{tabular}{ |c|c|cc|cc|cc|cc|cc|cc|c|} 
\hline
\multirow{2}{*}{N} &Optimal & \multicolumn{13}{c|}{Distributed Algorithm } \\
\cline{3-15}
&Solution& \multicolumn{2}{c|}{$\rho=4$ }&   \multicolumn{2}{c|}{$\rho=5$} &   \multicolumn{2}{c|}{$\rho=6$} &   \multicolumn{2}{c|}{$\rho=7$} &   \multicolumn{2}{c|}{$\rho=8$} &   \multicolumn{2}{c|}{$\rho=9$} &  speedup\\
\hline
7 & 55.7&47.7&-14.0\%&48.4&\textbf{-13.0\%}&44.2&-21.0\%&39.0&-30.0\%&29.3&-47.0\%&18.5&-67.0\% & 317\\
\hline
8 & 66.3&58.1&-12.0\%&59.4&\textbf{-10.0\%}&58.8&-11.0\%&55.0&-17.0\%&46.5&-30.0\%&29.0&-56.0\% & 1360\\
\hline
9 & 75.2&68.1&-9.0\%&68.2&\textbf{-9.0\%}&66.2&-12.0\%&61.7&-18.0\%&49.9&-34.0\%&32.8&-56.0\% & 6217\\
\hline
10 & 84.4&76.5&\textbf{-9.0\%}&75.9&-10.0\%&74.3&-12.0\%&71.1&-16.0\%&60.1&-29.0\%&37.5&-56.0\% & 30229\\
\hline
11 & 93.5&85.5&-9.0\%&86.5&-8.0\%&86.6&\textbf{-7.0\%}&83.3&-11.0\%&77.4&-17.0\%&47.7&-49.0\% & 139808\\
\hline
12 & 104.0&92.7&-11.0\%&92.9&-11.0\%&93.8&\textbf{-10.0\%}&92.0&-12.0\%&79.3&-24.0\%&50.5&-51.0\% & 730198\\
\hline
\end{tabular}
\end{adjustbox}
\caption{Average of utility values versus $\rho$ for distributed algorithm versus $N$ and $\rho$. First column presents the optimal value of joint problem. The distributed algorithm's mean values are provided in the 1st entry of each $\rho$ column, the relative mean utility loss compared to optimal solution in the 2nd entry of the $\rho$ column. The last column provides
the speedup of the distributed algorithm over the exhaustive search.}
\label{Table2}
\end{center}
\end{table*}

To benchmark the results of the distributed algorithm versus the optimal solution obtained by exhaustive search, 
we provide an overview of the simulation results for $N \in \{7,8,\cdots,12\}$ in Table \ref{Table2}. 
The table's second column provides the mean utility of the optimal solution for each value of $N$.
For each value of $\rho$, a double column lists the mean utility of the distributed algorithm (left) and its difference to the optimal mean utility (right). 
The best values for a given $N$ are highlighted. For $N=7,8,9$, the distributed algorithm's best mean utility results from $\rho=5$, for $N=10$ from $\rho=4$, and for 
$N=11, 12$ from $\rho=6$. The last column of Table~\ref{Table2} shows the computation time speed-up of the distributed algorithm over the exhaustive search, e.g., for $N=12$ the distributed algorithm is approximately $10^6 $ times faster than the exhaustive search. 
This emphasizes the scalability and efficiency of our distributed approach compared to finding the joint problem's optimal solution.

\section{Conclusion} \label{conclusions}

The goal of this paper is to provide a general framework for initiating a distributed machine learning (DL) algorithm by UEs in future wireless networks, 
where the aggregation nodes of  DL algorithm are among UEs, referred to as leaders which can be fit to 5G and 6G in a straightforward manner. 
We introduce two new indices that express the willingness of each UE to be a leader (LII) and 
the willingness of other UEs to select it as leader (LXI) without sharing any internal and external  state of UEs to any entity. 
We have shown how LIIs and LXIs can simplify the formulation of the UE clustering problem, called \textit{joint leader selection and follower association} problem, where 
followers are UEs who update ML models and share their result with leaders selected among UEs. 
We introduced related constraints and solutions for the joint problem and the distributed algorithm to enhance scalability. 
Through simulation results we have demonstrated that the distributed algorithm's mean utility approaches the optimum solution with a gap of  $\approx 
10\%$ at average. 
This gap is tolerable considering the benefits of the distributed algorithm: lower complexity by multiple orders of magnitude compared to finding an optimal solution, 
and scalability of a decentralized cluster formation process. Moreover, the distributed algorithm is also superior to any potential joint problem solved by a central server 
as it protects private and sensitive data, e.g., the trust levels between UEs.

\bibliographystyle{IEEEtran}

\bibliography{myref}

\begin{thebibliography}{10}
\providecommand{\url}[1]{#1}
\csname url@samestyle\endcsname
\providecommand{\newblock}{\relax}
\providecommand{\bibinfo}[2]{#2}
\providecommand{\BIBentrySTDinterwordspacing}{\spaceskip=0pt\relax}
\providecommand{\BIBentryALTinterwordstretchfactor}{4}
\providecommand{\BIBentryALTinterwordspacing}{\spaceskip=\fontdimen2\font plus
\BIBentryALTinterwordstretchfactor\fontdimen3\font minus
  \fontdimen4\font\relax}
\providecommand{\BIBforeignlanguage}[2]{{%
\expandafter\ifx\csname l@#1\endcsname\relax
\typeout{** WARNING: IEEEtran.bst: No hyphenation pattern has been}%
\typeout{** loaded for the language `#1'. Using the pattern for}%
\typeout{** the default language instead.}%
\else
\language=\csname l@#1\endcsname
\fi
#2}}
\providecommand{\BIBdecl}{\relax}
\BIBdecl

\bibitem{10056683}
S.~Parsaeefard and A.~Leon-Garcia, ``Efficient transfer learning in {6G},'' in
  \emph{2022 IEEE Future Networks World Forum (FNWF)}, 2022, pp. 314--319.

\bibitem{10258360}
Q.~Duan, J.~Huang, S.~Hu, R.~Deng, Z.~Lu, and S.~Yu, ``Combining federated
  learning and edge computing toward ubiquitous intelligence in 6g network:
  Challenges, recent advances, and future directions,'' \emph{IEEE
  Communications Surveys \& Tutorials}, vol.~25, no.~4, pp. 2892--2950, 2023.

\bibitem{9352556}
C.~Yang, Y.-M. Cheung, J.~Ding, and K.~C. Tan, ``Concept drift-tolerant
  transfer learning in dynamic environments,'' \emph{IEEE Transactions on
  Neural Networks and Learning Systems}, vol.~33, no.~8, pp. 3857--3871, 2022.

\bibitem{5288526}
S.~J. {Pan} and Q.~{Yang}, ``A survey on transfer learning,'' \emph{IEEE
  Transactions on Knowledge and Data Engineering}, vol.~22, no.~10, pp.
  1345--1359, Oct 2010.

\bibitem{mcmahan2017communication}
B.~McMahan, E.~Moore, D.~Ramage, S.~Hampson, and B.~A. y~Arcas,
  ``Communication-efficient learning of deep networks from decentralized
  data,'' in \emph{Artificial intelligence and statistics}.\hskip 1em plus
  0.5em minus 0.4em\relax PMLR, 2017, pp. 1273--1282.

\bibitem{HFL}
M.~S.~H. Abad, E.~Ozfatura, D.~GUndUz, and O.~Ercetin, ``Hierarchical federated
  learning across heterogeneous cellular networks,'' in \emph{ICASSP 2020 -
  2020 IEEE International Conference on Acoustics, Speech and Signal Processing
  (ICASSP)}, 2020, pp. 8866--8870.

\bibitem{FSM}
D.-J. Han, M.~Choi, J.~Park, and J.~Moon, ``Fedmes: Speeding up federated
  learning with multiple edge servers,'' \emph{IEEE Journal on Selected Areas
  in Communications}, vol.~39, no.~12, pp. 3870--3885, 2021.

\bibitem{MAFL}
Z.~Qu, X.~Li, J.~Xu, B.~Tang, Z.~Lu, and Y.~Liu, ``On the convergence of
  multi-server federated learning with overlapping area,'' \emph{IEEE
  Transactions on Mobile Computing}, vol.~22, no.~11, pp. 6647--6662, 2023.

\bibitem{10423299}
T.~Zhang, K.-Y. Lam, and J.~Zhao, ``Device scheduling and assignment in
  hierarchical federated learning for internet of things,'' \emph{IEEE Internet
  of Things Journal}, pp. 1--1, 2024.

\bibitem{lalitha2018fully}
A.~Lalitha, S.~Shekhar, T.~Javidi, and F.~Koushanfar, ``Fully decentralized
  federated learning,'' in \emph{Third workshop on bayesian deep learning
  (NeurIPS)}, vol.~2, 2018.

\bibitem{gholami2022trusted}
A.~Gholami, N.~Torkzaban, and J.~S. Baras, ``Trusted decentralized federated
  learning,'' in \emph{2022 IEEE 19th Annual Consumer Communications \&
  Networking Conference (CCNC)}.\hskip 1em plus 0.5em minus 0.4em\relax IEEE,
  2022, pp. 1--6.

\bibitem{liu2022decentralized}
W.~Liu, L.~Chen, and W.~Zhang, ``Decentralized federated learning: Balancing
  communication and computing costs,'' \emph{IEEE Transactions on Signal and
  Information Processing over Networks}, vol.~8, pp. 131--143, 2022.

\bibitem{cattrysse1992survey}
D.~G. Cattrysse and L.~N. Van~Wassenhove, ``A survey of algorithms for the
  generalized assignment problem,'' \emph{European journal of operational
  research}, vol.~60, no.~3, pp. 260--272, 1992.

\bibitem{kunalpaper}
T.~Moscibroda and R.~Wattenhofer, ``{Facility location: distributed
  approximation},'' \emph{24th ACM Symposium on the Principles of Distributed
  Computing}, Las Vegas, Nevada, USA, July 2005.

\bibitem{9207469}
C.~Briggs, Z.~Fan, and P.~Andras, ``Federated learning with hierarchical
  clustering of local updates to improve training on non-iid data,'' in
  \emph{2020 International Joint Conference on Neural Networks (IJCNN)}, 2020,
  pp. 1--9.

\bibitem{long2023multi}
G.~Long, M.~Xie, T.~Shen, T.~Zhou, X.~Wang, and J.~Jiang, ``Multi-center
  federated learning: clients clustering for better personalization,''
  \emph{World Wide Web}, vol.~26, no.~1, pp. 481--500, 2023.

\bibitem{D2.2}
\BIBentryALTinterwordspacing
B.~Priyanto, G.~Berardinelli, M.~H. Ramlov, B.~Coll-Perales, L.~Lusvarghi,
  K.~Aghababaiyan, J.~G. Sempere, M.~S. Ribes, N.~Kusashima, B.~Raaf,
  S.~Roessel, P.~M. de~Sant~Ana, H.~Klessig, M.~Li, and F.~Foukalas,
  ``Deliverable {D}2.2: Refined definition of scenarios, use cases and service
  requirements for in-x subnetworks,'' \emph{Project 6G-SHINE}, 2024. [Online].
  Available: \url{https://6gshine.eu/deliverables-ii}
\BIBentrySTDinterwordspacing

\bibitem{zookeeper}
{ZooKeeper Leader Election}, ``{ Available:
  \url{https://zookeeper.apache.org/doc/r3.8.4/recipes.html}},'' \emph{Last
  Published:}, May 2024.

\bibitem{convexboyd}
S.~Boyd and L.~Vandenberghe, \emph{Convex Optimization}.\hskip 1em plus 0.5em
  minus 0.4em\relax Cambridge, U.K., 2004.

\bibitem{saeedehbook}
S.~Parsaeefard, A.~R. Sharafat, and N.~Mokari, \emph{Robust Resource Allocation
  in Future Wireless Networks}.\hskip 1em plus 0.5em minus 0.4em\relax
  Springer, 2017.

\bibitem{Knuth_2Notes}
D.~E. Knuth, ``Two notes on notation,'' \emph{Amer. Math. Monthly}, vol.~99,
  no.~5, pp. 403 -- 422, 1992.

\end{thebibliography}







\end{document}